\documentclass[twocolumn,showpacs,preprintnumbers,amsmath,amssymb,superscriptaddress,aps]{revtex4}

\usepackage{graphicx}
\usepackage{dcolumn}
\usepackage{bm}
\usepackage{natbib}

\begin{document}

\newcommand{\width}{0.45}
\title{Coherence Properties of Individual Femtosecond Pulses
of an X-ray Free-Electron Laser}

\author{I. A. Vartanyants}
\altaffiliation[Corresponding author: ]{Ivan.Vartaniants@desy.de}
\affiliation{Deutsches Elektronen-Synchrotron DESY, Notkestr. 85, D-22607 Hamburg, Germany}
\affiliation{National Research Nuclear University, "MEPhI", 115409 Moscow, Russia}
\author{ A. Singer }
\affiliation{Deutsches Elektronen-Synchrotron DESY, Notkestr. 85, D-22607 Hamburg, Germany}
\author{ A. P. Mancuso}
\altaffiliation[Experimental Realization: ]{Adrian.Mancuso@xfel.eu}
\altaffiliation[Present address: ]{European XFEL GmbH, Albert-Einstein-Ring 19, 22761 Hamburg, Germany}
\affiliation{Deutsches Elektronen-Synchrotron DESY, Notkestr. 85, D-22607 Hamburg, Germany}
\author{ O. Yefanov}
\affiliation{Deutsches Elektronen-Synchrotron DESY, Notkestr. 85, D-22607 Hamburg, Germany}
\author{ A. Sakdinawat}
\affiliation{University of California, Berkeley, CA 94720, USA}
\author{ Y. Liu}
\affiliation{University of California, Berkeley, CA 94720, USA}
\author{ E. Bang}
\affiliation{University of California, Berkeley, CA 94720, USA}
\author{ G.J. Williams}
\affiliation{SLAC National Accelerator Laboratory, 2575 Sand Hill Road, Menlo Park, CA 94025-7015, USA}
\author{ G. Cadenazzi}
\affiliation{ARC Centre of Excellence for Coherent X-ray Science, School of Physics, The University of Melbourne, Victoria, 3010, Australia}
\author{ B. Abbey}
\affiliation{ARC Centre of Excellence for Coherent X-ray Science, School of Physics, The University of Melbourne, Victoria, 3010, Australia}
\author{ H. Sinn}
\affiliation{European XFEL GmbH, Albert-Einstein-Ring 19, 22761 Hamburg, Germany}
\author{ D. Attwood}
\affiliation{University of California, Berkeley, CA 94720, USA}
\author{ K. A. Nugent}
\affiliation{ARC Centre of Excellence for Coherent X-ray Science, School of Physics, The University of Melbourne, Victoria, 3010, Australia}
\author{ E. Weckert}
\affiliation{Deutsches Elektronen-Synchrotron DESY, Notkestr. 85, D-22607 Hamburg, Germany}
\author{ T. Wang}
\affiliation{SLAC National Accelerator Laboratory, 2575 Sand Hill Road, Menlo Park, CA 94025-7015, USA}
\author{ D. Zhu}
\affiliation{SLAC National Accelerator Laboratory, 2575 Sand Hill Road, Menlo Park, CA 94025-7015, USA}
\author{ B. Wu}
\affiliation{SLAC National Accelerator Laboratory, 2575 Sand Hill Road, Menlo Park, CA 94025-7015, USA}
\author{ C. Graves}
\affiliation{SLAC National Accelerator Laboratory, 2575 Sand Hill Road, Menlo Park, CA 94025-7015, USA}
\author{ A. Scherz}
\affiliation{SLAC National Accelerator Laboratory, 2575 Sand Hill Road, Menlo Park, CA 94025-7015, USA}
\author{ J. J. Turner}
\affiliation{SLAC National Accelerator Laboratory, 2575 Sand Hill Road, Menlo Park, CA 94025-7015, USA}
\author{ W. F. Schlotter}
\affiliation{SLAC National Accelerator Laboratory, 2575 Sand Hill Road, Menlo Park, CA 94025-7015, USA}
\author{ M. Messerschmidt}
\affiliation{SLAC National Accelerator Laboratory, 2575 Sand Hill Road, Menlo Park, CA 94025-7015, USA}
\author{ J. L\"uning}
\affiliation{Laboratoire de Chimie Physique, Université Pierre et Marie Curie (Paris VI), Paris, France}
\author{ Y. Acremann}
\affiliation{ETH Z\"urich, Laboratorium f\"ur Festk\"orperphysik, Schafmattstr. 16, 8093 Z\"urich}
\author{ P. Heimann}
\affiliation{Advanced Light Source, Lawrence Berkeley National Laboratory, Berkeley, California 94720, USA}
\author{ D. C. Mancini}
\affiliation{Advanced Photon Source and Center for Nanoscale Materials, Argonne National Laboratory, Argonne, IL 60439}
\author{ V. Joshi}
\affiliation{Advanced Photon Source and Center for Nanoscale Materials, Argonne National Laboratory, Argonne, IL 60439}
\author{ J. Krzywinski}
\affiliation{SLAC National Accelerator Laboratory, 2575 Sand Hill Road, Menlo Park, CA 94025-7015, USA}
\author{ R. Soufli}
\affiliation{Lawrence Livermore National Laboratory, 7000 East Avenue, Mail Stop L-211, Livermore, California 94551, USA}
\author{ M. Fernandez-Perea}
\affiliation{Lawrence Livermore National Laboratory, 7000 East Avenue, Mail Stop L-211, Livermore, California 94551, USA}
\author{ S. Hau-Riege}
\affiliation{Lawrence Livermore National Laboratory, 7000 East Avenue, Mail Stop L-211, Livermore, California 94551, USA}
\author{ A.G. Peele}
\affiliation{ARC Centre of Excellence for Coherent X-ray Science  Department of Physics, La Trobe University, Melbourne, Victoria 3086, Australia}
\author{ Y. Feng}
\affiliation{SLAC National Accelerator Laboratory, 2575 Sand Hill Road, Menlo Park, CA 94025-7015, USA}
\author{ O. Krupin}
\affiliation{SLAC National Accelerator Laboratory, 2575 Sand Hill Road, Menlo Park, CA 94025-7015, USA}
\affiliation{European XFEL GmbH, Albert-Einstein-Ring 19, 22761 Hamburg, Germany}
\author{ S. Moeller}
\affiliation{SLAC National Accelerator Laboratory, 2575 Sand Hill Road, Menlo Park, CA 94025-7015, USA}
\author{ W. Wurth}
\affiliation{Institut f\"ur Experimentalphysik and CFEL, University of Hamburg, Luruper Chaussee 149, 22761 Hamburg, Germany}
\date{\today}

\begin{abstract}
  Measurements of the spatial and temporal coherence of single, femtosecond x-ray pulses generated by the first hard x-ray free-electron laser (FEL), the Linac Coherent Light Source (LCLS), are presented. Single shot measurements were performed at 780 eV x-ray photon energy using apertures containing double pinholes in "diffract and destroy" mode. We determined a coherence length of $17~\mu$m  in the vertical direction, which is approximately the size of the focused LCLS beam in the same direction. The analysis of the diffraction patterns produced by the pinholes with the largest separation yields an estimate of the temporal coherence time of 0.6 fs. We find that the total degree of transverse coherence is 56\% and that the x-ray pulses are adequately described by two transverse coherent modes in each direction. This leads us to the conclusion that 78\% of the total power is contained in the dominant mode.
\end{abstract}

\pacs{41.60.Cr, 42.55.Tv, 42.25.Kb, 42.25.Hz}

\maketitle
Coherence is the fundamental property of light waves produced by laser sources. In combination with ultrashort pulses, they yield insight into basic questions in physics through real-time observation and control of atomic scale structure and dynamics \cite{10}. The recent development of x-ray free-electron lasers (XFEL) in the extreme ultraviolet (XUV) \cite{11} and the hard x-ray range \cite{9,XFEL1,XFEL2} with their unprecedented peak brightness, short pulse duration -- below 10 fs -- and importantly, a high degree of transverse coherence, open new frontiers in the study of the structure and dynamics of matter.

The intense, coherent and ultra-short x-ray pulses produced by XFELs promise important new insights in biology \cite{1,2} condensed matter physics \cite{3} and atomic physics \cite{4}. They have already paved the way for new approaches to protein crystallography using nanocrystals \cite{5} and the imaging of single viruses \cite{6}.
Spatial coherence is essential for applications such as coherent x-ray diffractive imaging (CXDI) \cite{13,14,15}, x-ray holography \cite{16} and x-ray photon correlation spectroscopy (XPCS) \cite{17}. The recovery of structural information from coherent imaging experiments relies on a high degree of spatial coherence in the incident field to enable the phasing of the diffraction pattern \cite{7,8} produced by its scattering from the sample. Despite this importance no direct measurements of the coherence properties of XFEL beams from hard x-ray FELs have been reported, although estimates of the coherence properties of these sources have been available through simulations \cite{18,19}. Here we present measurements of the coherence of the Linac Coherent Light Source (LCLS) x-ray beam.

One of the most widely used methods for characterization of coherence is Young's experiment \cite{20}, where two small pinholes separated by a certain distance are illuminated. The visibility of the resultant interference pattern is a measure of the correlation within the wavefield incident at the two pinholes--that is, the transverse coherence of the illuminating beam. An analysis of the contrast of these interference fringes, as a function of their distance from the center of the diffraction pattern, can also yield a measurement of the temporal coherence.  This method has been successfully employed with light beams \cite{21}, x-rays at synchrotrons \cite{22} and in the XUV energy range also using pulsed sources \cite{23,24,25,26,27}.

The goal of this experiment is to characterize the coherence properties of individual, focused LCLS pulses using double pinhole apertures. These single shot measurements were performed in the ``diffract-and-destroy'' mode \cite{15}.

\begin{figure}[bt!]
\includegraphics[width=\width\textwidth]{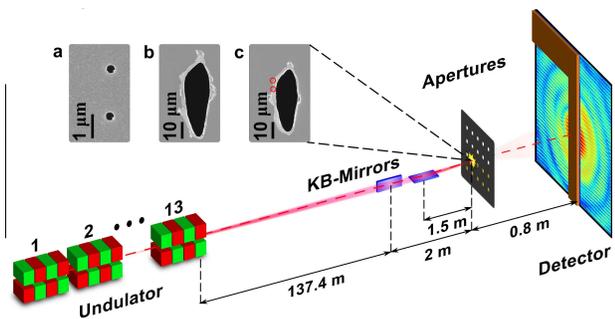}
\caption{(Color online)
A sketch of the experiment showing thirteen undulator modules, a set of KB-mirrors focusing the beam on a sample frame and the detector, protected from the direct beam by a beamstop. The inset shows SEM images of different apertures before (a) and after (b) the exposure of a single LCLS pulse. The inset (c) shows a case when the center of the beam missed the double pinhole (marked by the red circles).}
\end{figure}
The experiment was conducted at the soft x-ray research (SXR) instrument of the LCLS. A sketch of the experiment is shown in Fig. 1. The LCLS was operated with an electron bunch charge of 250 pC and with 13 undulator segments tuned to deliver 780 eV ($\lambda$=1.6 nm) x-ray photons. Under these conditions, LCLS is expected to reach its saturation regime \cite{9,12,19}. The duration of a single pulse was about 300 fs, determined from electron bunch measurements. The average energy was about 1 mJ per pulse.
The beam was delivered to the end-station through a beam transport system that includes three plane distribution mirrors and a monochromator comprised of a spherical mirror followed by a plane grating. The measurements presented here were performed with the monochromator grating replaced by a plane mirror. The limiting vertical aperture of the beam delivery system was twice the full-width at half maximum (FWHM) of the beam size at the grating at 800 eV. At the sample position, the beam was focused to a size of $5.7\pm0.4~\mu$m (FWHM) in the horizontal and $17.3 \pm2.4~\mu$m (FWHM) in the vertical direction (see Appendix \ref{ap:focus}) by a pair of bendable Kirkpatrick-Baez (KB) mirrors consisting of a silicon substrate coated with a 37.4 nm thick boron carbide reflective coating \cite{31,32,33}, with focal lengths of 1.5 m (V) and 2 m (H).

A multiple aperture array (see Appendix \ref{ap:apertures}) with varying pinhole separations in the range from 2 $\mu$m to 15 $\mu$m was positioned in the focus of the beam inside the Resonant Coherent Imaging (RCI) end-station (Fig. 1). After each shot on the sample, the array was moved to an unexposed sample position. To accumulate statistics, each pinhole configuration was measured several times giving 110 patterns in total. Interference patterns were recorded by a Princeton Instruments PI-MTE 2048B direct illumination Charge Coupled Device (CCD) with 2048$\times$2048 pixels, each 13.5$\times$13.5 $\mu$m$^2$ in size, positioned 80 cm downstream from the apertures (Fig. 1). A 3 mm wide rectangular beamstop manufactured from B$_4$C was positioned in front of the CCD to protect it from exposure to the direct FEL beam.

\begin{figure}[bt!]
\includegraphics[width=\width\textwidth]{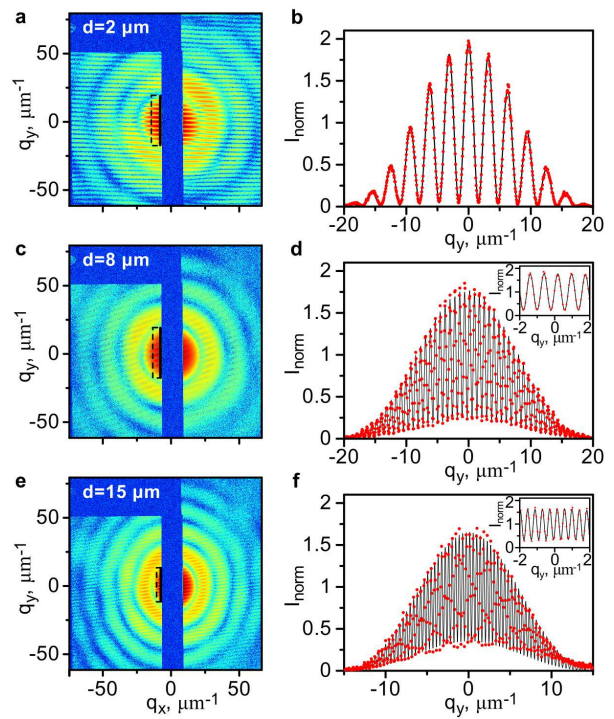}
\caption{(Color online)
Measured diffraction patterns. Left column: Interference fringes from pinholes separated by 2 $\mu$m (a), 8 $\mu$m (c) and 15 $\mu$m (e) each exposed to a single shot of the LCLS beam as a function of the transverse momentum transfer $q_x$, $q_y$. The area used for the analysis of the transverse coherence is shown by the dashed black rectangle. Right column: Results of the fit (black lines), to the experimental data (red dots) shown in the left column. Line scans marked with the black solid lines in the left column are presented. The insets in (d) and (f) show enlarged regions.
}
\end{figure}
Figure 2 shows typical single-shot diffraction patterns measured with different pinhole separations. For small separations between the pinholes a high contrast diffraction pattern was observed implying a high degree of coherence on that length scale. For larger separations the visibility of the fringes is slightly reduced due to the partial coherence of the incoming beam.

The interference pattern, $I(P)$, observed in a double pinhole experiment at the point $P$ of the detector for narrow-bandwidth radiation can be described by the following expression \cite{20,28,footnote1}
%
%
\begin{equation}
  \begin{split}
    I(P)=I_0(P)\left(1+\left|\gamma_{12}^{eff}(\tau)\right|\cos\left[\omega\tau-\alpha_{12}(\tau)\right]  \right),
\end{split}
\end{equation}
where $I_0(P)$ is the Airy distribution due to diffraction through a round pinhole of diameter $D$ \cite{20}, $\tau$ is the time delay for the radiation to reach point $P$ from different pinholes, $\omega$ is a mean frequency of the incoming radiation. The modulus of the effective complex degree of coherence $|\gamma_{12}^{eff}(\tau)|$ in equation (1) is defined as $|\gamma_{12}^{eff}(\tau)|= 2\left[\sqrt{I_1I_2}/(I_1+I_2)\right]|\gamma_{12}(\tau)|$, where $\gamma_{12}(\tau)$ is the intrinsic complex degree of coherence, $I_{1,2}$ are intensities incident at the pinholes one and two and $\alpha_{12}(\tau)$ is the relative phase. When the incident intensities at both pinholes are identical, $|\gamma_{12}^{eff}(\tau)|= |\gamma_{12}(\tau)|$.

The analysis of the diffraction data was performed by fitting expression (1) to each measured diffraction pattern (see Fig. 2 and Appendix \ref{ap:analysis}). In this analysis, we considered a region of the diffraction pattern shown in Fig. 2 where $|\gamma_{12}^{eff}(\tau)|\approx|\gamma_{12}^{eff}(0)|$ and $\alpha_{12}(\tau)\approx\alpha_{12}(0)$ are good approximations as the time delay associated with the path-length difference is much shorter than the coherence time $\tau_c$, $\tau\ll\tau_c$ (see below). The modulus of the effective complex degree of coherence $|\gamma_{12}^{eff}|$ at a particular pinhole separation was determined for each shot (Fig. 3). A Gaussian fit through the 'best' shots (those that provided the highest degree of coherence and shown as black squares in Fig. 3) gives an upper estimate for the root-mean-square (rms) value of the transverse coherence length, $l_y^c=16.8\pm1.7~\mu$m, of the focused LCLS beam in the vertical direction.

Our analysis shows a significant variation of the effective degree of coherence between different pulses for the same pinhole separation (see Fig. 3). While this variation could be explained by shot-to-shot fluctuations of the coherence properties of the XFEL beam, it may also arise from uncertainty in the position of the incoming beam with respect to the center of the pinhole pair, which leads to a difference in intensity at each pinhole. The value of the effective complex degree of coherence, $|\gamma_{12}^{eff}|$, can be significantly lower than the intrinsic complex degree of coherence, $|\gamma_{12}|$, if these incident intensities are not equal. To determine the possible maximum deviation of the incident pulses with respect to the center of the pinhole pair we observed that some pulses were not centered on the apertures and did not destroy the pinholes (see inset (c) in Fig. 1). We analyzed SEM images of these apertures and determined a maximum deviation of 11 $\mu$m in the vertical direction. 
The impact of this positional uncertainty on the contrast, deduced from the beam size and the Gaussian fit through the 'best' shots, is described by the blue dashed line in Fig. 3. Our data indicate that most of the experimentally determined values lie in the range between the two lines corresponding to the 'best' and the maximum offset shots. From this we conclude that this positional uncertainty is the dominant cause of the apparent shot-to-shot variation of the complex degree of coherence.

\begin{figure}[bt!]
\includegraphics[width=\width\textwidth]{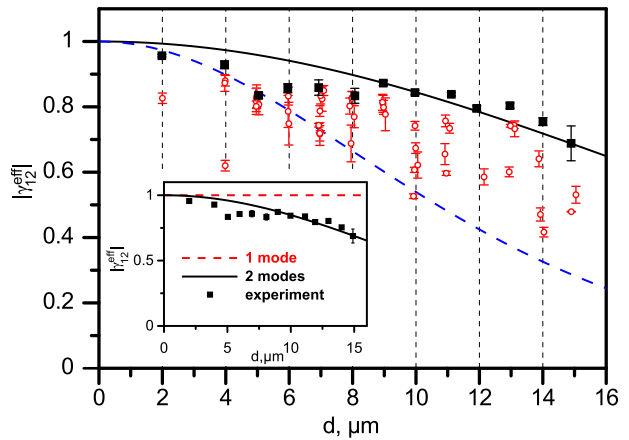}
\caption{(Color online)
The modulus of the effective complex degree of coherence, $|\gamma_{12}^{eff}|$, as a function of pinhole separation. The experimental values determined from the fitting procedure are shown by red circles. The error bars show the standard deviation of these values. A Gaussian function (black line) has been fit to the best shot values (black squares) which gives a coherence length of  16.8$\pm$1.7 $\mu$m. The blue dashed line shows the decrease in the value of $|\gamma_{12}^{eff}|$ due to the maximum measured offset between the position of the apertures and the incident beam. (inset) The contribution of higher order modes to the modulus of the complex degree of coherence. The fully coherent case (single mode) is shown by the red dashed line. The two mode contribution is shown by the black line.
}
\end{figure}

Some of the pulses that illuminated apertures with larger pinhole separations (greater than 10 $\mu$m) were extremely bright. This allows the determination of the fringe visibility up to the edge of the detector, which corresponds to time delays of $\tau\approx0.6$ fs. At these conditions the time dependence in Eq. (1) was taken into account explicitly, providing a measurement of the temporal coherence for individual femtosecond pulses.  An average over six single pulses yields a temporal coherence time, $\tau_c\approx0.55\pm0.12$ fs (rms). For a beam with a Gaussian spectrum this value is in good agreement with the estimate \cite{20} $\tau_c\sim1/\sigma_\omega=0.4$ fs, where $\sigma_\omega=2.5 fs^{-1}$ is the rms bandwidth of the LCLS beam at that energy \cite{12}.

The knowledge of the transverse coherence length of the LCLS beam and an estimate of its vertical focus size is sufficient to determine the degree of transverse coherence, $\zeta_y$, in the vertical direction. We used the following definition of the total degree of transverse coherence \cite{30}
$\zeta=\int |W(\mathbf r_1,\mathbf r_2;\omega)|^2 \mbox d\mathbf r_1\mbox d\mathbf r_2\cdot\left( \int I(\mathbf r)\mbox d\mathbf r \right)^{-2} $, where $W(\mathbf r_1,\mathbf r_2;\omega)$ is the cross spectral density (CSD) \cite{28}.
In the frame of the Gaussian Schell-model (GSM) \cite{28,30} the total CSD, $W(\mathbf r_1,\mathbf r_2;\omega)$, factorizes into the product of two independent components \footnote{For brevity, we omit $\omega$ from future equations}, $W(\mathbf r_1,\mathbf r_2)= W(x_1,x_2) W(y_1,y_2)$. The same holds for the intensities, $I(\mathbf r)= I(x) I(y)$. As a result the total degree of transverse coherence can be presented as a product of the horizontal and vertical contributions $\zeta=\zeta_x\cdot\zeta_y$, where $\zeta_{x,y}=(l^c_{x,y}/\sigma_{x,y})\cdot\left[ (l^c_{x,y}/\sigma_{x,y})^2+4 \right]^{-1/2}$ and $l^c_{x,y},~\sigma_{x,y}$ are the transverse coherence lengths and the beam sizes (rms), respectively. For the focused LCLS beam we determined $\zeta_y=0.75\pm0.08$. A similar value, $\zeta_x$, is expected in the horizontal direction as the source size and the beam divergence at FEL sources have comparable magnitudes in both directions \cite{24}. Thus the total degree of coherence for the full beam is $\zeta=0.56\pm0.12$. This is comparable with the value obtained in simulations \cite{19} for similar LCLS parameters.

The properties of highly coherent beams, such as XFEL beams, can also be conveniently described by their mode decomposition \cite{28}. 
The CSD, $W(y_1,y_2)$, can be decomposed into a sum of independent modes $W(y_1,y_2)=\sum_j \beta_jE_j^*(y_1)E_j(y_2)$, where $\beta_j$ is the contribution of each mode $E_j(y)$. Applying the same GSM model to the 'best' shots in the vertical direction yields  $\beta_1/\beta_0=0.14\pm0.05$ and $\beta_2/\beta_0=0.02\pm0.01$ for the first and for the second mode, respectively. This indicates that for separations of up to 15 $\mu$m, which corresponds to the FWHM of the beam, two modes are sufficient to describe the coherence properties of the beam in the vertical direction (see inset in Fig. 3).

Using the mode decomposition of the CSD the intensity in each direction can be described by $I(x)=\sum_j\beta_j^xI_j(x)$, where $I_j(x)$ is the normalised intensity distribution of the $j$-th mode. The total power of the wavefield $P=\int I(x)I(y) \mbox dx\mbox dy$ is determined in this case by
$P=P_0+P_{0,1}+P_{1,0}+\cdots=\beta_0^x\beta_0^y+\beta_0^x\beta_1^y+\beta_1^x\beta_0^y+\cdots$, where we have neglected the contribution of modes higher than two. From this expression, the relative power of the dominant mode is $P_0/P=\left[1+\beta_1^y/\beta_0^y+\beta_1^x/\beta_0^x \right]^{-1}$.
Extrapolating our results to the horizontal direction we estimate that $78\pm8$\% of the total FEL beam power is concentrated in the dominant ``TEM$_{00}$'' mode. This value is substantially higher than at any existing x-ray source at that wavelength (it is about 1\% for synchrotron sources \cite{30}).

Using the same model, the photon beam emittance $\varepsilon_y$ in the vertical direction is given by \cite{30} $\varepsilon_y=\sigma_y\cdot\sigma_y'=\lambda/(4\pi\zeta_y)$, where $\sigma_y$ is the rms of the source size and $\sigma_y'$ is the rms divergence of the photon beam. Substituting into this expression the measured value of the degree of transverse coherence, $\zeta_y$, we find that the emittance of the LCLS beam is $\varepsilon_y=0.17\pm0.02$ nm rad. This agrees well with typical values reported for the LCLS photon beam at 800 eV with the source size $\sigma_y=20~\mu$m and divergence $\sigma_y'=8.5~\mu$rad \cite{12}. For a diffraction limited beam with $\zeta_y=1$, the same source size and x-ray photon energy would have a smaller divergence of about 6.4 $\mu$rad (see Fig. A in Appendix \ref{ap:divergence}).

In conclusion, we have measured the coherence properties of the LCLS using the focused x-ray beam at a photon energy of 780 eV. The total degree of transverse coherence was found to be 56\%, from which we estimate that 78\% of the total power is contained in the dominant mode. Furthermore, the temporal coherence of the LCLS beam was measured to be 0.6 fs, in good agreement with an averaged LCLS spectrum at these energies. We foresee that this single shot methodology for the coherence measurement of high-power, pulsed x-ray sources developed here can be extended to investigate the performance of the LCLS in different conditions of operation. 
Finally, understanding the high degree of coherence at XFEL sources -- as demonstrated in this work -- provides a solid foundation for future coherence-based experiments that exploit these bright, coherent x-ray beams.	

Portions of this research were carried out on the SXR Instrument at the Linac Coherent Light Source (LCLS), a division of SLAC National Accelerator Laboratory and an Office of Science user facility operated by Stanford University for the U.S. Department of Energy. The SXR Instrument is funded by a consortium whose membership includes the LCLS, Stanford University through the Stanford Institute for Materials Energy Sciences (SIMES), Lawrence Berkeley National Laboratory (LBNL), University of Hamburg through the BMBF priority program FSP 301, and the Center for Free Electron Laser Science (CFEL). Use of the Center for Nanoscale Materials was supported by the U. S. Department of Energy, Office of Science, Office of Basic Energy Sciences, under Contract No. DE-AC02-06CH11357. Part of this work was performed under the auspices of the U.S. Department of Energy by Lawrence Livermore National Laboratory under Contract No. DE-AC52-07NA27344. Financial support by the CNRS through the PEPS SASLELX program is acknowledged by J. Luning. We acknowledge a careful reading of the manuscript and suggestions made by Z. Huang and P. Emma.
%

\bigskip
\bigskip
\begin{Large}
 \textsc{\qquad\qquad\quad~ Appendix}
\end{Large}
\section{Focus size measurements}
\label{ap:focus}
To estimate the average size of the focus, we exploited a shot-to-shot variation in alignment between the beam and the apertures due to instabilities in the beam position and the sample stage in the plane of the sample. Using the coordinates of undamaged pinholes and the corresponding scattered intensities measured as an integrated signal at the CCD we determined a few points on the tails of the intensity distribution curve at the position of the pinholes. Fitting a Gaussian through these points gives an average beam size of $17.3 \pm 2.4~\mu$m FWHM in the vertical direction and $5.7 \pm 0.4~\mu$m FWHM in the horizontal, in the plane of the apertures. 

Analysis of the highly offset shots in horizontal direction allowed us to estimate how uniform the coherence properties of the pulses are as a function of transverse position within the pulse. We compared the values of the complex degree of coherence for strongly horizontally offset and vertically centered pulses with the remainder of the pulses. These offset pulses also displayed high coherence, which implies that the coherence properties of the LCLS pulses appear to be spatially uniform.

\section{Apertures}
\label{ap:apertures}
The apertures were fabricated by electroplating a 1.3 $\mu$m thick gold layer on top of a 100 nm silicon nitride substrate supported by windows etched in a 200 $\mu$m thick silicon wafer. The 1.3 $\mu$m gold film attenuates the beam by eight orders of magnitude at the photon energies used here. The sample was 20 mm x 25 mm in size and consisted of 4 arrays of 11 x 13 windows, for a total of 572 windows. Each window, 50 x 50 $\mu$m$^2$ in size, contained a pair of pinholes. The distance between individual windows was 768 $\mu$m in both directions. The pinhole diameter varied from 340 nm for the smallest separation to 500 nm for the largest separation to account for the reduction in intensity due to the larger separations probing the less intense regions of the beam.

\section{Data Analysis}
\label{ap:analysis}
The analysis of diffraction patterns has shown that some contain a contribution from an incoherent background, mostly to one side, which leads to a reduced visibility of fringes. For the analysis considered here, regions with sufficient signal on the opposite side of the beamstop (marked with the dashed rectangle in Fig. 2 (a,c,e)) were considered. These were divided into vertical slices 10 pixels wide. The number of slices varied from 5 to 10 depending on the pinhole size. The single shot values of $|\gamma_{12}^{eff}|$, shown in Fig. 3, are each an average over these slices, with the error bars given by the statistical variation (standard deviation) between these slices. The diffraction patterns, where the incoherent background was present in the data in the analyzed region, were identified by a high variation of the fit parameters between different slices. These interference patterns, as well as the patterns with poor signal, were excluded from our evaluation. The following parameters were determined while fitting equation (1) to the experimental data:  the incident intensity $I_0^{in}$, the modulus of the effective complex degree of coherence $|\gamma_{12}^{eff}|$, the relative phase $\alpha_{12}(0)$ of the wavefield between the pinholes, the pinhole separation $d$, the pinhole diameter $D$, and the position of the beam centre $q_{x,y}^0$.
The small inclination angle of about 30 mrad in the vertical positioning of the pinholes was taken into account during the data analysis.
Measured values of $|\gamma_{12}^{eff}|$ were corrected for the finite width of the modulation transfer function (MTF) of the detector. The detector MTF was measured independently by observing the variation of the contrast produced by two pinholes at a fixed pinhole separation as a function of the sample to detector distance using the beamline 13-3 at the SSRL synchrotron source. A 25 fringes/mm (rms) Gaussian MTF function for our detector was determined.

\section{Divergence of the beam}
\label{ap:divergence}
\begin{minipage}[h]{8.1 cm}
\includegraphics[width=1\textwidth]{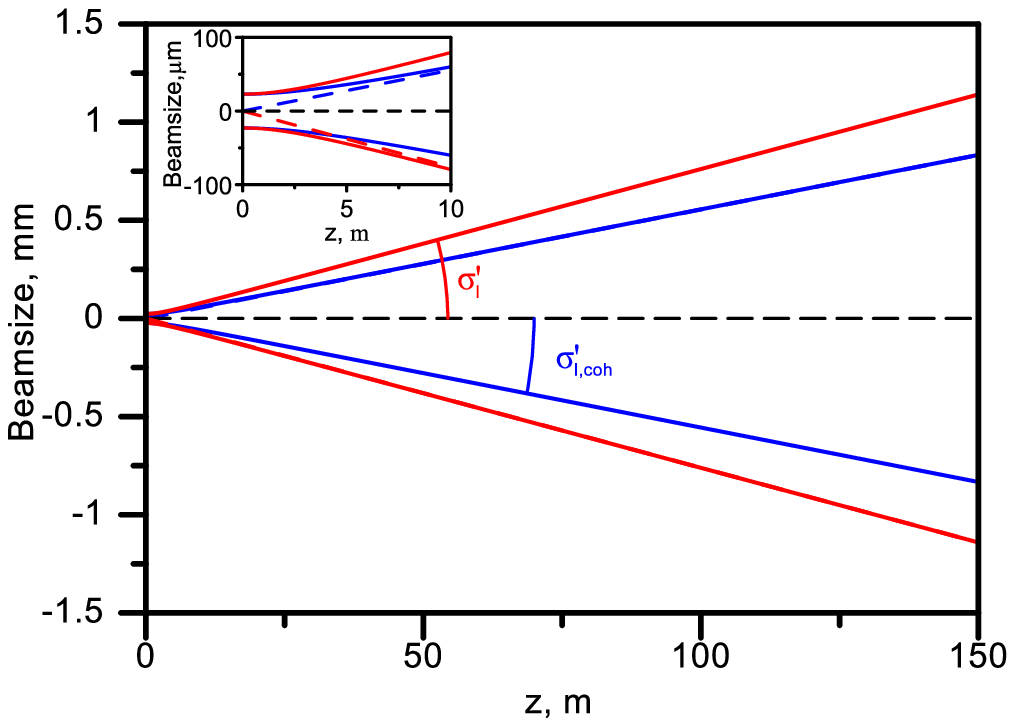}
\begin{small}
  \textsc{FIG.~A:~}(Color online)
Divergence of the LCLS beam. Simulations for the partially coherent LCLS beam with the degree of coherence 75 \% (red line) (source size $\sigma=20\mu$m and divergence $\sigma'=8.5 \mu$rad) compared to a fully coherent (diffraction limited) source with the same source size  and divergence $\sigma'_{coh}= 6.4 \mu$rad (blue line) as a function of the propagation distance from the source. The inset shows an enlarged region.
\end{small}
\end{minipage}


\begin{thebibliography}{99}
\bibitem{footnote1} The natural bandwidth of the LCLS radiation is less than 1\% at soft X-ray energies. Equation (1) is obtained under the paraxial and far-field approximations. It is  assumed that diffraction patterns from different pinholes completely overlap at the detector position. 
  Expression (1) also assumes that the radiation is stationary.
  With duration of the LCLS pulses, $T_0=300$ fs, being much longer than the estimated coherence time $\tau_c\approx0.6$ fs ($T_0\gg\tau_c$) one may safely use stationary conditions.
\end{thebibliography}
\end{document}